\documentclass[aps,prx,twocolumn,superscriptaddress, nofootinbib,  longbibliography]{revtex4-2}
\usepackage{graphicx} % needed for figs
\usepackage{dcolumn}  % needed for some tables
\usepackage{bm}    % for math
\usepackage{amssymb}  % for math
\usepackage{amsfonts, amsmath, amssymb, mathtools}
\usepackage{amsmath}

\usepackage{lipsum}
\usepackage[normalem]{ulem}
\usepackage{commath}
\usepackage{xcolor}
\usepackage{hyperref}

\usepackage{scalerel}[2016-12-29]

\usepackage[binary-units=true]{siunitx}
\usepackage{hyperref}
\usepackage{cleveref}
\RequirePackage{textcase}
\DeclareUnicodeCharacter{2009}{\,} 

\begin{document}

\raggedbottom

\title{On-demand transposition across light-matter interaction regimes in bosonic cQED}

\author{Fernando~Valadares}
\email[Corresponding author: ]{fernando.valadares@u.nus.edu}
\author{Ni-Ni~Huang}
\affiliation{Centre for Quantum Technologies, National University of Singapore, Singapore}
\author{Kyle~Chu}
\affiliation{Centre for Quantum Technologies, National University of Singapore, Singapore}
\affiliation{Horizon Quantum Computing, Singapore}
\author{Aleksandr~Dorogov}
\affiliation{Centre for Quantum Technologies, National University of Singapore, Singapore}
\author{Weipin~Chua}
\affiliation{Department of Physics, National University of Singapore, Singapore}
\author{Lingda~Kong}
\affiliation{Centre for Quantum Technologies, National University of Singapore, Singapore}
\author{Pengtao~Song}
\affiliation{Centre for Quantum Technologies, National University of Singapore, Singapore}
\author{Yvonne~Y.~Gao}
\email[Corresponding author: ]{yvonne.gao@nus.edu.sg}
\affiliation{Centre for Quantum Technologies, National University of Singapore, Singapore}
\affiliation{Department of Physics, National University of Singapore, Singapore}
\date{\today}

\begin{abstract}
The diverse applications of light-matter interactions in science and technology stem from the qualitatively distinct ways these interactions manifest, prompting the development of physical platforms that can interchange between regimes on demand. Bosonic cQED employs the light field of high-Q superconducting cavities coupled to non-linear circuit elements, harnessing the rich dynamics of their interaction for quantum information processing. However, implementing fast switching of the interaction regime without deteriorating the cavity coherence is a significant challenge. We present the first experiment to achieve this feat, combining nanosecond-scale frequency tunability of a transmon coupled to a cavity with lifetime of hundreds of microseconds. Our implementation affords a range of new capabilities for quantum information processing; from fast creation of cavity Fock states using resonant interaction and interchanging tomography techniques at qualitatively distinct interaction regimes on the fly, to the suppression of unwanted cavity-transmon dynamics during idle evolution. By bringing flux tunability into the bosonic cQED toolkit, our work opens up a new paradigm to probe the full range of light-matter interaction dynamics within a single platform and provides valuable new pathways towards robust and versatile quantum information processing.

\end{abstract}

\maketitle

% british spelling + authors comments

\begin{figure}[htp!]
\includegraphics[scale=1]{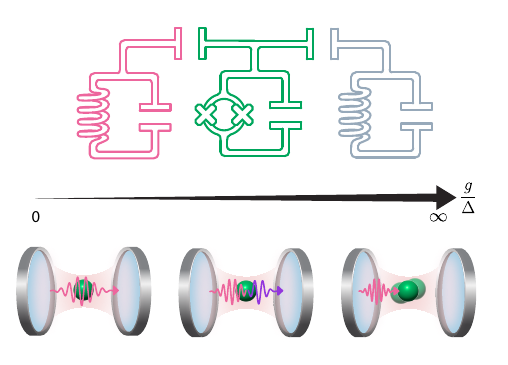 }
\caption{\textbf{Distinct regimes of light-matter interactions.}
The dynamics of light and matter depend on their frequency detuning $\Delta$ and interaction strength $g$. When $\Delta \ll g$, the systems are resonant and directly exchange energy. When $\Delta \gg g$, they are decoupled. In between these limits is the dispersive coupling regime, in which the energy of one system is shifted by excitations in the other. We realize this model in bosonic cQED with a long-lived cavity acting as the light field (pink) coupled to a transmon with tunable frequency (green) representing matter. A second low-Q resonator (grey) is used to measure the transmon state.
} 
\label{fig:light-matter} 
\end{figure}

The interaction of light and matter is at the origin of numerous phenomena, from photosynthesis to photovoltaics, from the fundamental structures of atoms to quantum information processing. This ubiquitous physical concept is critical to understand nature and develop new technologies. The coupled dynamics of light and matter span several qualitatively distinct regimes, from a direct swap of energy when both systems are resonant to a full decoupling when their frequencies differ significantly (Fig.~\ref{fig:light-matter}). Each regime has unique properties and advantages. A platform capable of harnessing all of this rich physics - and interchanging between regimes on demand - will grant powerful new insights into both fundamental science and novel applications.

Many physical platforms originate from the interactions between light and matter~\cite{haroche_nobel_2013, buzek_cavity_1997, bruzewicz_trapped-ion_2019, albrecht_coupling_2013}. In particular, circuit quantum electrodynamics (cQED) stands out for its high engineerability and versatility~\cite{blais_circuit_2021}. In this system, the light field is built from a superconducting cavity, while nonlinear circuits based on Josephson junctions emulate the discrete energy spectrum of an atom~\cite{koch_charge-insensitive_2007}. Owing to the rapid advances in cQED hardware, it is now possible to build long-lived cavities and harness their rich dynamics through the manipulation of nonlinear elements such as transmons~\cite{joshi_quantum_2021,copetudo_shaping_2023}. This architecture, known as bosonic cQED, provides an ideal playground for exploring the qualitatively distinct behaviours of light-matter interactions, and a valuable platform for quantum error correction~\cite{sivak_real-time_2023, ofek_extending_2016, ni_beating_2023}, analogue simulations~\cite{wang_efficient_2020, braumuller_analog_2017} and metrology~\cite{wang_heisenberg-limited_2019}, and even for the long-term goal of engineering a universal quantum computer~\cite{, copetudo_shaping_2023}.
% we should cite for each thing to be consistent 

\begin{figure*}[hpt!]
\includegraphics[scale=1]{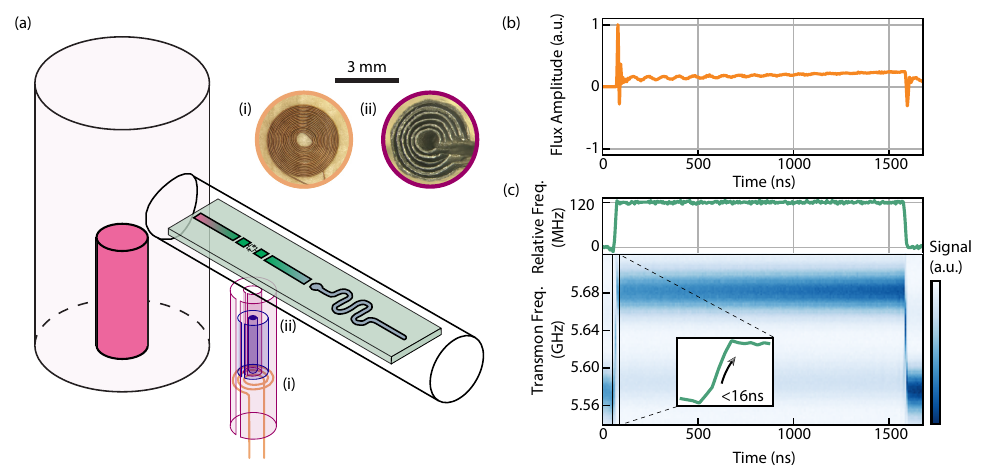}
\caption{\textbf{Fast-flux incorporation to bosonic cQED.} (a) Our hardware consists of a coaxial stub and a transversal coaxline~\cite{axline_architecture_2016} hosting a SQUID-based transmon and a readout resonator. The structure is made with high-purity aluminium and mounted in a dilution refrigerator below 10\,mK. The magnetic hose~\cite{sm} (not in proportion) is composed of concentric aluminium (light purple) and mu-metal (blue) layers and is inserted perpendicularly to the coaxline, aligned with the SQUID loop. Insets show the microscope pictures of (i) the source coil and (ii) the hose end closest to the transmon. (b) Example of corrected step flux pulse generated at the FPGA. (c) Corresponding frequency response of the transmon. Inset resolves the rising edge, which is measured to be shorter than 16\,ns.
}
\label{fig:device_and_flux}
\end{figure*}

While flux-controlled non-linear coupling elements between cavities have been demonstrated recently~\cite{chapman_high--off-ratio_2023, lu_high-fidelity_2023}, bosonic cQED still lacks the capability to harness the full range of light-matter interaction regimes on demand. Although transmons can readily be made flux-tunable with the use of a SQUID loop~\cite{koch_charge-insensitive_2007}, the addition of a strong, broadband flux line necessary to change the interaction dynamics within the coherence times of the system can critically reduce the cavity lifetime~\cite{reed_entanglement_2013, hofheinz_synthesizing_2009}. Because of this limitation, the community has so far focused operations in a single regime, which can be tailored for a specific experimental protocol but performs sub-optimally in other aspects~\cite{ofek_extending_2016, campagne-ibarcq_quantum_2020}. A device that combines real-time tunable transmons to high-Q cavities would therefore pave the way to further optimisation of existing techniques and the exploration of new quantum information processing protocols~\cite{sharma_quantum_2016, strauch_arbitrary_2010, strauch_all-resonant_2012, terhal_encoding_2016, terhal_towards_2020, aoki_control_2023, he_tunable_2019}.

In this paper, we demonstrate on-demand transposition across several distinct interaction regimes of a transmon and a long-lived cavity at nanosecond timescales. This is achieved by engineering a metamaterial structure that can efficiently route magnetic flux to a tunable transmon~\cite{navau_long-distance_2014, gargiulo_fast_2021} while preserving cavity coherences two orders of magnitude above that of the transmon. We showcase these features by mixing distinct coupling regimes within experiments that address central questions in quantum information processing. We prepare non-Gaussian quantum states in the cavity in a vacuum Rabi experiment via resonant coupling with the transmon. We then perform tomography on the prepared states using both Wigner and characteristic functions by tuning the cavity-transmon interaction to the strong and weak dispersive regimes, respectively. Finally, we demonstrate the suppression of undesired nonlinear distortions and the mitigation of cavity decoherence from transmon interference by decoupling the two elements. Our work significantly expands the functionalities of bosonic cQED architectures and provides the means for more sophisticated applications of light-matter interactions. 

\begin{figure*}[ht!]
\includegraphics[scale=1]{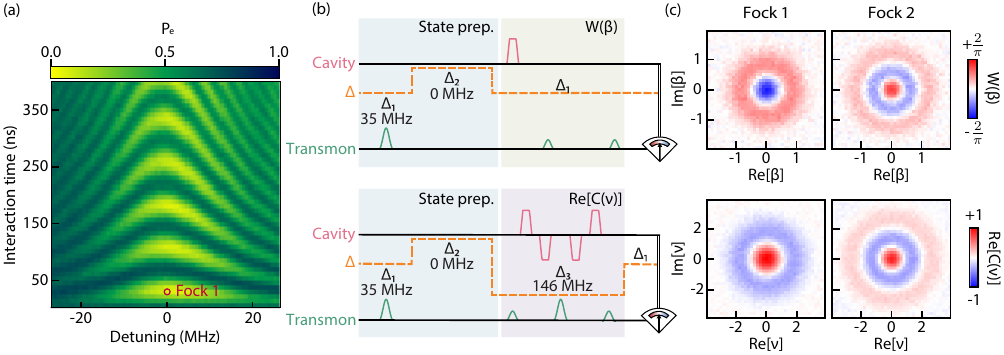}
\caption{\textbf{Preparation and tomography of non-Gaussian states.} (a) Vacuum Rabi oscillations between cavity and transmon as a function of interaction time and detuning. Fock states $|{1}\rangle$ and $|{2}\rangle$ are prepared at resonance through direct exchange of excitation with the transmon. (b) Diagrams showing Fock state preparation and tomography protocols using dynamical tuning of the transmon. The transmon is prepared in $|{e}\rangle$ at $\Delta_1/2\pi=35$\,MHz, then put into resonance with the cavity. After the interaction, it is either returned to $\Delta_1$ for Wigner tomography $W(\beta)$ (top) or to $\Delta_3/2\pi=146$\,MHz to measure the real part of the characteristic function $C(\nu)$ (bottom). $\pi$ ($\frac{\pi}{2}$)-pulse is denoted by the large (small) transmon pulse. (c) Tomography of Fock states $|{1}\rangle$ and $|{2}\rangle$ by measuring the Wigner function (top) and real part of the characteristic function (bottom). }
\label{fig:vacuum_rabi}
\end{figure*}

We achieve fast tunability of the transmon frequency while preserving high cavity coherence times by incorporating a carefully designed magnetic hose~\cite{navau_long-distance_2014, gargiulo_fast_2021} in a standard bosonic cQED device (Fig.~\ref{fig:device_and_flux}a). The system Hamiltonian is given by 
\begin{equation}
\label{eq:hamiltonian}
    \frac{\mathbf{\hat{H}}}{\hbar} = \omega_c \mathbf{\hat{a}}^\dag \mathbf{\hat{a}} + \omega_t \mathbf{\hat{b}}^\dag \mathbf{\hat{b}} + g\left(\mathbf{\hat{a}}^\dag \mathbf{\hat{b}} + \mathbf{\hat{a}} \mathbf{\hat{b}}^\dag\right) - \frac{\alpha}{2} \mathbf{\hat{b}}^\dag \mathbf{\hat{b}}^\dag \mathbf{\hat{b}} \mathbf{\hat{b}},
\end{equation}
where $\mathbf{\hat{a}}$ ($\mathbf{\hat{b}}$) is the cavity (transmon) mode annihilation operator, $\omega_{c,t}$ are the angular frequencies of the bare cavity and transmon, $g$ is the capacitive coupling factor between the circuits and $\alpha$ is the transmon anharmonicity. The tunability manifests as a change in the transmon-cavity detuning, $\Delta(t) = \omega_c - \omega_t\left(t\right)$ in response to the magnetic field provided by the hose.

The hose is an 8\,mm-long cylinder made of 11 alternating layers of mu-metal and superconducting aluminium, with respective thicknesses of 0.15\,mm and 0.10\,mm, assembled around a 1\,mm-diameter mu-metal core~\cite{sm}. The layers are designed to not close on themselves around the cylinder, keeping a 0.5\,mm gap along the hose axis to avoid superconducting loops. Due to the magnetic properties of this structure, the hose transfers magnetic fields with high efficiency, allowing the use of smaller, lower-impedance coils as the source. The source is made by winding 21 turns of superconducting wire into a planar coil with a maximum diameter of 3.4\,mm (see inset in Fig.~\ref{fig:device_and_flux}a). We double the magnetic field of the coil by stacking two such layers together and then attach them to one end of the hose. With this configuration, we can deliver a magnetic field of $80.6$\,nT/mA to the area of 1000\,$\mu \mathrm{m}^2$ forming the SQUID loop of the transmon.

To balance the efficient transfer of the magnetic field with the preservation of coherence times, we optimise the design of the bosonic cQED package, the transmon circuit and the magnetic hose itself. The lossy mu-metal layers of the hose are cut short 5\,mm away from the end closest to the chip to minimize leakage of the cavity and transmon energy. The transmon pads are made short to prevent its field from spreading out towards the hose. The hose is then placed as far as possible from the cavity to prevent its energy from leaking out. Since the hose must be aligned with the SQUID loop, the transmon is also further separated from the cavity, reducing their coupling. Their interaction is recovered by placing a superconducting strip between both circuits that allows their fields to travel along the chip (Fig.~\ref{fig:device_and_flux}a), reaching a coupling factor of $g/2\pi=6.65$\,MHz. A similar strip is placed between the transmon and the readout resonator to isolate the latter from the hose. Due to these design considerations, the cavity achieves average lifetimes of~200\,$\mu$s, comparable to standard non-tunable bosonic cQED architectures found in the literature. 

The effective execution of these delicate design choices of the hose and the coil results in full flux tunability of the transmon frequency from a maximum of $\omega_t^{\mathrm{max}}/2\pi~=~6.409$\,GHz down to values close to zero. The cavity frequency is at $\omega_c/2\pi = 5.740$\,GHz, so the transmon can access the resonant regime and also reach the decoupled regime at detuning $\Delta\gg g$. The small dimensions of the coil allow fast response to the drive. By connecting it to the digital-to-analogue converter port of an FPGA, we can send fast flux pulses to the coil and tune the transmon over several hundreds of MHz at nanosecond timescales.

Reliably switching between cavity-transmon interaction regimes requires precise control of the transmon frequency. However, the flux pulse generated by the room-temperature electronics is distorted by impedances present in the wiring, causing the transmon frequency to deviate from the target trajectory. These distortions mainly come from the inductance of the coil and the RF port of the bias tee, and can be reverted using the method described in Ref.~\cite{butscher_shaping_2018, sm}. We first characterize the distorted frequency response to a step flux pulse using a pi-scope experiment, consisting of repeated transmon spectroscopies at different points of the trajectory. The current $I_c(t)$ at the coil is obtained by inverting the frequency response of a symmetric SQUID transmon $\omega_t(t) \approx \left(\omega_t^{\mathrm{max}} + \alpha\right)\sqrt{|\cos(\pi kI_c/ \Phi_0)|} - \alpha$~\cite{koch_charge-insensitive_2007}, where $\Phi_0$ is the flux quantum and $\alpha/2\pi \approx 200$\,MHz. Here, $k = 0.039$\,$\Phi_0$/mA is a proportionality constant between the current and the magnetic flux threading the SQUID loop. From the $I_c(t)$ data, we digitally revert the distortions by training 11 first-order infinite-impulse response (IIR) and 2 finite-impulse response (FIR) filters. These filters are applied to a target flux pulse to produce the intended frequency trajectory. The corrected step pulse is shown in Fig.~\ref{fig:device_and_flux}b, and the resulting frequency response reliably reproduces a step function as shown in Fig.~\ref{fig:device_and_flux}c. The transmon frequency is stable over 1.5\,$\mu$s up to a variation of $\pm$0.2\%, and the rising edge has a duration below $16$\,ns. This switching time is much faster than the timescale of $2\pi/g=150$\,ns, allowing nonadiabatic control of the interaction.

We use this fast and stable frequency control to demonstrate the creation of non-Gaussian cavity states through vacuum Rabi oscillations (Fig.~\ref{fig:vacuum_rabi}a). The transmon, initially detuned from the cavity, is excited from the ground state $|{g}\rangle$ to the next energy level $|{e}\rangle$. Then, a step flux pulse tunes the transmon nonadiabatically to the cavity frequency, at which point they start to coherently swap energy. After the interaction, the transmon is measured. If it is found in $|{g}\rangle$, the missing energy is assumed to be in the cavity, preparing Fock state $|{1}\rangle$. Fig.~\ref{fig:vacuum_rabi}a shows the chevron pattern of the vacuum Rabi experiment for variable interaction time and detuning. This result can only be obtained if the transmon tuning is fast compared to its coupling to the cavity; otherwise, the transmon state would transform adiabatically into joint eigenstates of the whole system, which evolve trivially and do not show exchange of energy. State $|{1}\rangle$ is prepared in resonance after an interaction time $\tau_{\mathrm{int}}\approx30$\,ns, while $|{2}\rangle$ is prepared by repeating the same protocol twice with total $\tau_{\mathrm{int}}\approx50$\,ns. The state preparation is much shorter than the relevant coherence times of the system and avoids the use of numerically-optimised pulses that are often necessary to create complex cavity states~\cite{ khaneja_optimal_2005, heeres_implementing_2017, eickbusch_fast_2022}. This experiment provides the basis for more complex state preparation and control techniques in the resonant regime~\cite{hofheinz_synthesizing_2009, strauch_all-resonant_2012}.

To further showcase the versatility of our implementation, we characterize the Fock states with both Wigner and characteristic function tomographies by changing the transmon and cavity detuning. The ability to choose between these two protocols is an advantage since each gives ready access to different observables. In Wigner tomography, we directly probe the photon number parity of the bosonic state, a critical parameter in error correction protocols~\cite{michael_new_2016, mirrahimi_dynamically_2014}, and is typically optimal for systems in the strong dispersive regime. The characteristic function, on the other hand, measures the Fourier transform of the Wigner function and offers a more intuitive picture to study certain effects such as photon loss~\cite{pan_protecting_2023}. Contrary to Wigner, this tomography is usually done in the weak dispersive regime to decrease the nonlinear distortions during the protocol. The two techniques also have different resilience to errors. For example, the characteristic function protocol uses echo conditional displacement (ECD) gates to cancel out the effects of low-frequency noise~\cite{eickbusch_fast_2022}. We measure the prepared $|{1}\rangle$ and $|{2}\rangle$ states using both tomographies by tuning the transmon to either strong or weak dispersive regimes (Fig.~\ref{fig:vacuum_rabi}c). The results show the high quality of the state preparation and that our device can reliably perform tomography at different interaction regimes on demand. The ability to tune Hamiltonian parameters on the fly is thus a powerful tool for tailoring the tomography protocol to the requirements of any experiment.

\begin{figure}[htp!]
\includegraphics[scale=1]{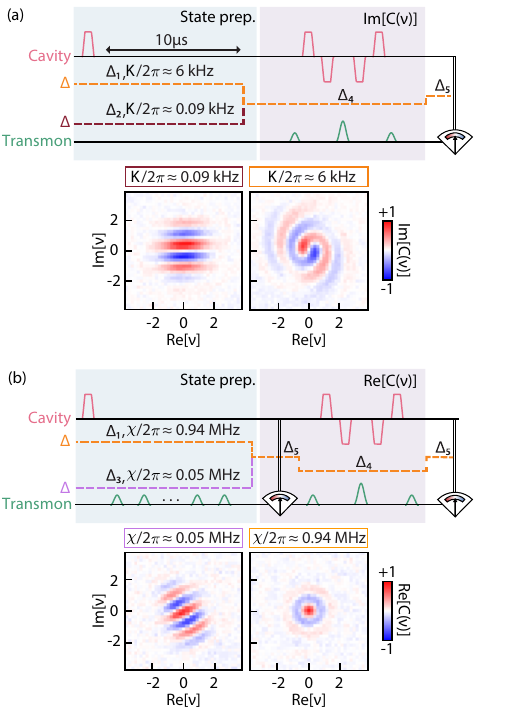}
\caption{\textbf{Suppression of undesired cavity dynamics.} (a) Mitigation of cavity self-Kerr. The cavity is prepared at coherent state $|{\alpha = 2.5}\rangle$ and allowed to evolve for 10\,$\mu$s at $\Delta_1/2\pi = 63$\,MHz ($K_1/2\pi=6$\,kHz) and $\Delta_2/2\pi=196$\,MHz ($K_2/2\pi=0.09$\,kHz). The imaginary part of the characteristic function $C(\nu)$ is measured at the end, showing that the state is virtually undisturbed with $K_2$, but significantly distorted during evolution with $K_1$. (b) Decoupling between transmon and cavity by turning off the dispersive interaction. The protocol is similar to the previous experiment, but the transmon is excited by a sequence of 10 $\frac{\pi}{2}$-pulses, each followed by an interval of 400\,ns. The real part of the characteristic function of the final state shows that the transmon excitation and decoherence cause the complete loss of phase information from the cavity state at $\chi/2\pi=0.94$\,MHz. In contrast, the coherent state still preserves its phase after evolution at $\chi/2\pi\approx0.05$\,MHz ($\Delta_3/2\pi =596$\,MHz). In both experiments, the tomography and the transmon readout are performed at detunings of $\Delta_4/2\pi = 146$\,MHz and $\Delta_5/2\pi = 101$\,MHz. }
\label{fig:kerr_chi}
\end{figure}

Beyond the creation of non-Gaussian cavity states and flexible tomography, another critical requirement of quantum information processing is the protection of the quantum state during idle evolution. In the dispersive regime ($g\ll \Delta$), the dynamics of the cavity are governed by $-\chi \mathbf{\hat{a}}^\dag \mathbf{\hat{a}} \mathbf{\hat{b}}^\dag \mathbf{\hat{b}} - \frac{K}{2} \mathbf{\hat{a}}^\dag \mathbf{\hat{a}}^\dag \mathbf{\hat{a}} \mathbf{\hat{a}}$, where $\chi$ is the dispersive shift between cavity and transmon and $K$ is the self-Kerr of the cavity. Although the self-Kerr is typically small, these extra dynamics can quickly distort the cavity state, motivating the search for ways to control and mitigate nonlinearities using drives or alternative circuits~\cite{zhang_engineering_2019, sivak_kerr-free_2019, chapman_high--off-ratio_2023}. Using fast-flux tunability, we can suppress the self-Kerr effect in the bosonic state by simply decoupling it from the transmon (Fig.~\ref{fig:kerr_chi}a). We demonstrate this suppression with the evolution of a coherent state $|{\alpha}\rangle$ with an average photon number of $|\alpha|^2\approx6.25$. We allow the state to evolve freely for 10~$\mu$s at two different flux points with $K/2\pi\approx0.09$~kHz and $K/2\pi\approx6$~kHz. Details for every flux point used are given Ref.~\cite{sm}. After this interval, we tune the transmon frequency to measure the imaginary part of the characteristic function. The tomography reveals a stark contrast between the cavity dynamics at each point as shown in Fig.~\ref{fig:kerr_chi}a. At the high self-Kerr point, the state is strongly distorted, losing the well-defined phase of a coherent state. In comparison, the evolution at low self-Kerr keeps the coherent state virtually unchanged, with its properties preserved due to the suppressed cavity-transmon interactions. 

The same strategy is used to tune down the dispersive interaction and mitigate the propagation of transmon errors to the bosonic state. We simulate a scenario where fluctuations in the transmon energy cause dephasing in the cavity (Fig.~\ref{fig:kerr_chi}b). Here, the transmon is initialized in $|{g}\rangle$ and the cavity is prepared in $|{\alpha}\rangle$. We apply a pulse to put the transmon in a superposition $\left(|{g}\rangle + |{e}\rangle\right)/\sqrt{2}$ and allow the system to evolve for $\tau=$~400\,ns. Due to the dispersive shift $\chi$ of the cavity frequency when the transmon is excited, the system forms an entangled state  $|{ g,~\alpha}\rangle + |{e,~\alpha e^{i\chi\tau}}\rangle $ (up to a normalization factor), which continuously decoheres into a statistical mixture of $|{ g,~\alpha}\rangle$ and $|{e,~\alpha e^{i\chi\tau}}\rangle$ due to transmon decoherence. This procedure is repeated 10 times, inducing a random walk of the phase of $\alpha$ with steps of $\chi\tau$. At the end of the procedure, the transmon is projected onto $|{g}\rangle$ and the real part of the characteristic function is measured. We compare the evolution of the cavity state for two different flux points. In the first case, the dispersive shift is strongly suppressed to $\chi/2\pi \approx 0.05$\,MHz. and the real characteristic function of the final state is shown in Fig.~\ref{fig:kerr_chi}b (left panel). Due to the decoupling, the coherent state still preserves a well-defined phase given by the direction of the fringes, with an overall rotation due to the residual $\chi$. In the second case, the dynamics of the cavity are strongly intertwined with the transmon at $\chi/2\pi=0.94$\,MHz, resulting in a much stronger distortion of the state. As such, the phase of the cavity state is completely scrambled and the fringes of the coherent state are fully washed out (Fig.~\ref{fig:kerr_chi}b).

These experiments show a high-Q bosonic cQED design that leverages the many regimes of light-matter interactions and combines them to create new functionalities. The resourcefulness of our solution is demonstrated in multiple cases of interest for quantum information processing. We perform vacuum Rabi oscillations by nonadiabatically switching the transmon and cavity into resonance and use this technique to prepare non-Gaussian cavity states without the need for numerically optimised pulses. The states are measured with both Wigner and characteristic function, respectively in the strong and weak dispersive regimes, showing flexibility in extracting information from the cavity. We then decouple the cavity and transmon to show how to `freeze' the evolution of the cavity state and protect it against spurious transmon dynamics during idle times.% through the suppression the action of self-Kerr nonlinearities and the unwanted dispersive shift interaction.

As a first proof-of-principle demonstration, there are certainly nonidealities in our system. For instance, the transmon decoherence time (few $\mu$s) in our hardware is noticeably lower compared with fixed-frequency transmons~\cite{gargiulo_fast_2021, yoshihara_decoherence_2006} as well as planar devices specifically optimised for flux noise insensitivity~\cite{braumuller_concentric_2016, hutchings_tunable_2017}. We expect significant improvements in future iterations that incorporate the known mitigating strategies such as alternative circuit designs~\cite{braumuller_concentric_2016, hutchings_tunable_2017} and further optimisation of the hose strength to reduce SQUID loop sizes. We believe that this work conclusively demonstrates the appeal and viability of our strategy, and will prompt future exploration towards more robust and tunable bosonic cQED hardware implementations.

Our results also hint at a much wider range of applications using bosonic cQED architectures equipped with fast-flux tunability. For example, resonant control can be expanded to synthesize arbitrary qudit states~\cite{hofheinz_synthesizing_2009} and enact individual transitions in the cavity spectrum~\cite{strauch_all-resonant_2012}. Dynamically decoupling the transmon can mitigate the cross-talk in multi-cavity quantum processor architectures. Controlling the idle cavity evolution can be used to optimise tomography protocols, which commonly suffer from coherent errors due to spurious transmon coupling during the measurement. Finally, the ability to put different light-matter dynamics in juxtaposition also opens up new possibilities for trotterising Hamiltonian for quantum simulation and computation~\cite{mezzacapo_digital_2014}.

Achieving an architecture capable of tuning the coherent interactions of light and matter on the fly has been a longstanding milestone across many quantum hardware platforms. Our implementation in bosonic cQED provides a highly promising strategy towards addressing this considerable engineering challenge of balancing good cavity lifetimes with fast on-demand flux tunability. Our results bring a set of powerful new functionalities into the repertoire of high-quality superconducting cavities that will stimulate the creation of new protocols for quantum information processing. Furthermore, with this newfound feature, the bosonic cQED platform can now unlock and more effectively harness the vast potential of the full range of dynamics in light-matter interaction.

\textbf{Acknowledgements}
We thank Prof. Gerhard Kirchmair, Dr.~Stefan Oleschko, and Dr.~Ian Yang for their valuable input on the initial development and construction of the magnetic hose. We thank Dr.~Guangqiang Liu and Prof.~Michel Devoret for providing the quantum amplifier. We thank Adrian Copetudo for the assistance in the chip fabrication. This research is supported by the Ministry of Education, Singapore, under grant ID T2EP50222-0017 and the National Research Foundation, Singapore, under grant ID NRFF12-2020-0063.  

\appendix
\section{Device Design \& System Parameters}

The device used in this work is based on standard bosonic cQED systems~\cite{blais_cavity_2004, girvin_circuit_2014}, consisting of one high-Q cavity, one SQUID transmon, and a low-Q resonator for readout. We simulate the electromagnetic fields of the device using Ansys finite-element High-Frequency Simulation Software (HFSS), and obtain the Hamiltonian parameters using the energy participation ratio (EPR) approach~\cite{minev_energy_2021}. The key system properties, such as the frequency of each circuit and the pair-wise non-linear couplings between them, are iteratively refined to meet the target parameters. In this section, we describe the details on the design considerations and resulting properties of the main elements in the device, namely, the cavity, the SQUID transmon, the magnetic hose, and the coil. 

\subsection{Hamiltonian parameter optimisation}
The design of Hamiltonian parameters takes into consideration the variable SQUID transmon frequency. Specifically, the transmon's design incorporates two critical frequency regions: one optimized for strong interaction with the cavity and a second region intentionally decoupled from the cavity. Even in this decoupled state, transmon maintains substantial coupling with the readout resonator. This dual-region design allows versatile functionality for effective cavity interactions and significant coupling for readout purposes. The first frequency region is used for cavity state preparation and tomography, while the second, designed for readout with minimized cross-Kerr effects, is later repurposed for cavity evolution under low nonlinearities, as illustrated in Fig.~4 in the main text.

\begin{figure}[hbt!]
\includegraphics[scale=1]{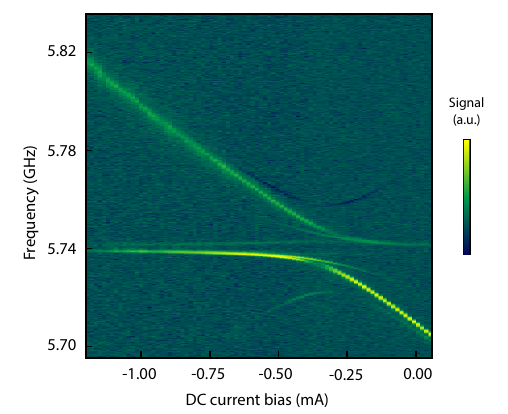}
\caption{\textbf{Vacuum Rabi avoided crossing.} As the transmon is tuned into to resonance with the cavity, anti-crossing with a gap $2g\approx 14.60$\,MHz is observed. This is consistent with the simulated target parameter of $g=7.44$\,MHz.}
\label{figs1:avoided_crossing}
\end{figure}

The transmon's maximum frequency is designed to be 6.5\,GHz, strategically positioned between the frequencies of the cavity (5.74\,GHz) and the readout resonator (6.9\,GHz). This specific configuration enables the transmon to achieve resonance with the cavity within the tunable range to perform the vacuum Rabi experiment. In addition, the transmon is also consistently operated below the cavity frequency to avoid entering the straddling regime~\cite{krantz_quantum_2019}, which ensures optimal performance and suppresses undesired interactions.

\begin{table*}[hbt]
\begin{ruledtabular}
\begin{tabular}{c|ccccc}
\textbf{Points} &  $\Delta/2\pi$ (MHz) & $\chi_{qc}^{\text{exp}} /2\pi$ (MHz) & $\omega_{t}^{\text{exp}}/2\pi$ (GHz) & $\omega_t^{\text{sim}} /2\pi$ (GHz) & $K^{\text{sim}}/2\pi$ (kHz)  \\ \hline
\textbf{A} &  35 & 1.67  & 5.705 & 5.696 & 44 \\
\hline
\textbf{B} & 63  & 0.94  & 5.677 & 5.662 & 6 \\ 
\hline
\textbf{C} & 101 & 0.57  & 5.639 & 5.635 & 1.9 \\
\hline
\textbf{D} & 146 & 0.29 & 5.594 & 5.547 & 0.19 \\ 
\hline
\textbf{E} & 196 & 0.18 & 5.544 & 5.492 & 0.09 \\
\hline
\textbf{F} & 596 &  0.05 & 5.144 & 5.214 & 0.005 \\
\end{tabular}
\end{ruledtabular}
\caption{\textbf{Summary of key system parameters at selected flux points.} The simulated and measured main parameters, organized in ascending order of detuning $\Delta$ from the cavity frequency, at different flux points relevant to the experiments reported in the main text. For values of the cavity self-Kerr, numbers are quoted based on simulation results only as some of the smaller self-Kerr values cannot be precisely measured due to experimental limitations.
\label{tab:table3}}
\label{tab:flux_point_info_chi}

\end{table*}

The cavity, transmon, and readout resonator are capacitively coupled, with the coupling strength determined by the overlap between their electromagnetic fields. In practice, the coupling factor $g$ between each of the elements mainly depends on geometrical parameters such as the sizes of the transmon capacitor pads and their distance from the cavity and the readout resonator. In our design, we chose a coupling factor of $g=7.44$\,MHz between the cavity and the transmon. This is sufficiently large for fast exchange of excitations between the two elements on the nanosecond timescale while allowing a safe distance between the magnetic hose and the cavity.

Experimentally, we observed consistent agreement between measured system parameters and design targets. For example, the vacuum Rabi splitting illustrated in Fig.~\ref{figs1:avoided_crossing} indicates $g=7.30$\,MHz. This reduced slightly to $g=6.65$\,MHz in subsequent thermal cycles, as measured from the rate of the vacuum Rabi oscillations shown in the main text (Fig.~2a). Similar agreement between experiment and simulation is observed across the full transmon frequency range. We summarised these parameters at selected flux points referenced in experiments reported in the the main text in Table~\ref{tab:flux_point_info_chi}.

\subsection{Quality factor of the cavity}
Apart from the Hamiltonian parameters, the coherence property of the cavity mode is also carefully considered in our design process. Our primary focus is minimizing cavity losses directly through the mu-metal layers of the magnetic hose. For this particular analysis, we deliberately exclude both the transmon and the readout resonator from the simulation to eliminate potential indirect losses associated with cavity hybridization with other circuits.

The simulations show that the critical factor to be optimized is the distance between the cavity and the hose. Notably, the Q-factor experiences a decrease of roughly an order of magnitude for each millimetre of approximation. Thus, making this spacing long enough is critical for our device. However, the hose has to be aligned with the SQUID loop at the middle of the transmon. Thus, distancing the hose from the cavity means the transmon must also be placed further away. 
    
Moreover, the transmon pads cannot be arbitrarily stretched to reach the cavity, as this would cause the transmon field to spread out of the chip and to couple to the loss channel of the hose. To enhance the cavity-transmon coupling, we employ a superconducting strip, approximately 2.85\,mm long, strategically placed on both sides of the transmon, as shown in Fig.~\ref{fig:hose_chip}a. We ensure that this strip is short enough that it does not introduce any resonant mode that is close in frequency to the other elements. The strips effectively guide the transmon field to the cavity and readout resonator modes, recovering their coupling without requiring them to be in close proximity.

In the finalised design, the center of the hose is placed approximately 
3.65\,mm away from the edge of the cavity, resulting in a simulated upper bound for the cavity lifetime of 1.6\,ms. Experimentally, we conducted standard cavity $T_1$ measurements across multiple thermal cycles, observing $T_1$ to range from 100\,$\mu$s to 250\,$\mu$s. This is mainly limited by the poor coherence times of the transmon ($T_1\approx 15\,\mu$s, $T_2\approx 0.5\text{-}4\,\mu$s). In addition, we also observed that the measured cavity $T_1$ does not depend on the flux applied to the coil, indicating that the introduction of the magnetic flux to the system is not the dominant loss in our device.  

\subsection{Magnetic Hose}
\begin{figure*}[ht!]
\includegraphics[scale=1]{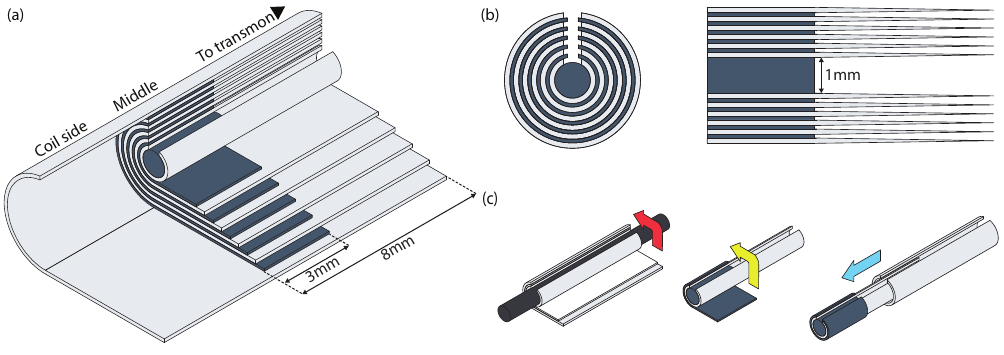}
\caption{\textbf{Assembly and design features of the hose} (a) Half-open view of assembled hose 3D model. Aluminium (mu-metal) layers are shown in light (dark) gray. The 3D model depicts the hose in three distinct sections: ``Coil side'',  where the coil is inserted; the ``Middle'' section, where the mu-metal is placed to ensure high magnetic field transferability and hold the inner aluminium layers; and the output, labeled as ``To transmon'', made exclusively of aluminium layers and functioning as a high-pass filter. (b) Cross-section and side cut of the hose showing adaptations to prevent superconducting loops. In the left picture, the mu-metal layers are shown to have a smaller cut than the aluminium layers, providing protection against aluminium-aluminium contact. In the right picture, the aluminium layers are shown to be made thin at the output end to make the spacing between layers larger. (c) Hose assembly. First step (red arrow): aluminium layers are bent into cylinders. The first layer is bent around a placeholder 1\,mm-diameter rod. Subsequent layers use paper (colored as white) as placeholders for mu-metal, matching the 0.1\,mm thickness. After all aluminium layers are bent, the placeholders are removed and the layers disassembled. Second step (yellow arrow): the first aluminium layer is assembled around the mu-metal core. The first mu-metal layer is bent around the structure. Third step (blue arrow): the next aluminium layer is assembled to the structure. The second and third steps are repeated with every layer until the hose is assembled.}
\label{fig: hose}
\end{figure*}

Conceptually, we can divide the hose into three sections. The input section, defined by the outer aluminium layer, allows for the insertion of the small coil which generates the magnetic field traveling through the hose. This section can be elongated until it protrudes out of the device, creating a surface to be clamped from the outside, as depicted in Fig.~\ref{fig: hose}a and labeled ``Coil side''. Next is the middle section, which houses aluminium, mu-metal layers, and the core wire, focusing on high transferability of the magnetic field due to the high magnetic permeability of the mu-metal. However, the properties of the mu-metal are very fragile, and its quality significantly deteriorates during the hose manufacturing. Consequently, this section is intentionally kept as short as possible, as illustrated in Fig.~\ref{fig: hose}a and labeled ``Middle''.

The output section, depicted in Fig.~\ref{fig: hose}a and labeled ``To transmon'', consists solely of aluminium layers, with a gap of 0.1\,mm between each of them. Functioning as a high-pass filter, this section aims to attenuate any electromagnetic field emanating from the output end of the hose. As the hose guides flux to the transmon, the filter serves to prevent the circuit's field from interacting with the lossy mu-metal layers. The length of this section is critical: if it is excessively long, the magnetic field provided to the transmon may be insufficient. Conversely, if it is too short, the filter may fail to effectively isolate the mu-metal from the transmon and cavity, thereby affecting their coherence times.

If there are any superconducting loops in the hose, the input magnetic field will generate stable supercurrents that oppose the field. Preventing superconducting loops requires careful adjustments to the hose so the aluminium layers don't come in contact with each other. Firstly, in the middle section, widening the mu-metal layers slightly beyond the aluminium layers helps prevent the outer layers from folding into the inner layers, as illustrated in Fig.~\ref{fig: hose}b. In the output section, using sandpaper to reduce the thickness of the aluminium layers, from 0.15\,mm to 0.06\,mm - 0.08\,mm, increases the space between layers, lowering the risk of contact. During the process of folding the aluminium layers into cylinders, it is critical to ensure that they are made as straight as possible to prevent any bending into each other at the output. These considerations collectively contribute to minimizing the possibility of unwanted superconducting loops.

The actual magnetic hose used in this device is designed and implemented based on the techniques outlined in Ref.~\cite{gargiulo_fast_2021} with the additional considerations outlined above. The procedure for making a hose starts with cutting five mu-metal rectangular pieces from a 0.1\,mm-thick sheet and six aluminium pieces from a 0.15\,mm-thick sheet, colored as dark and light gray, respectively, as shown in Fig.~\ref{fig: hose}a. Then, a mu-metal core is cut from a wire with a diameter of 1\,mm. These sheets are systematically layered around the core, with alternating aluminium and mu-metal layers. Precise calculations of the width of each layer ensure that each layer covers the perimeter of the underlying layer, leaving a 0.5\,mm gap to prevent superconducting loops.

All mu-metal pieces are cut to 3\,mm in length, striking the balance of being the shortest length to suppress losses while still being easily manageable for handling. The five innermost aluminium layers are 8\,mm in length. They fully cover the mu-metal layers, with a 3\,mm section overlap between them and the remaining 5\,mm extending into the vacuum. The outermost aluminium layer, elongated to 25\,mm, accommodates the insertion of a coil, as illustrated in the Fig.~2a(i) in the main text.

\subsection{Design and implementation of the coil}
\label{sec:coil}

\begin{figure}[hbt!]
\includegraphics[scale=1]{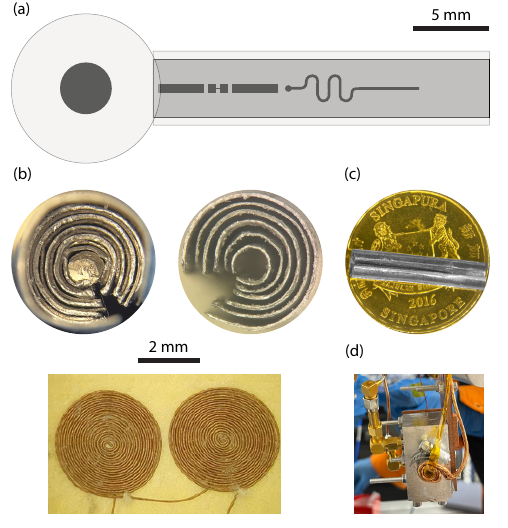}
\caption{\textbf{Full device illustration}. (a) The system illustrated from a top view, showing the coaxial stub cavity (left), transmon with the extended superconducting strips (middle), and meandering readout resonator (right). (b) Pictures of the magnetic hose and coil to scale. The left image shows the input side of the hose, where both aluminium and mu-metal layers can be seen. The right image displays the output end, where only aluminium layers are present. To the best of our knowledge, the aluminium layers don't come in contact with each other, and the 0.5\,mm cut remains open along the hose. The bottom picture shows the double coil used as field source. After being wound, the two layers are glued together with epoxy and assembled to the hose. (c) Example of a hose compared with a Singapore coin of a diameter of 21\,mm. (d) Picture of the device showing the thermalisation of the hose using copper braids.}
\label{fig:hose_chip}
\end{figure}

To ensure fast response to the input current, we need a relatively small coil. We crafted a disk-shaped coil (illustrated in the Fig.~2a(i) in the main text and Fig~\ref{fig:hose_chip}b) by winding 19 to 21 turns of a 0.05\,mm-diameter superconducting wire up to a maximum diameter of 3.4\,mm. Due to the small dimensions, the coil must be crafted under a microscope with the aid of tweezers. To enhance field strength, we used a double coil configuration. This comes at the cost of decreasing the coil bandwidth, but its response is still fast enough to enact nonadiabatic dynamics in the vacuum Rabi experiment. After winding, the two layers are stacked with the aid of epoxy and then affixed to a dielectric support. The support is also affixed to the outer layer of the hose to improve thermalisation. The coil lead wires are soldered to an SMA adaptor that connects to the RF flux line.  

\section{Wiring \& Measurement setup}
\subsection{Cryogenic configuration}

Our device is machined out of pure aluminium with a tunable transmon realised as a DC-SQUID featuring two double-angle evaporated Josephson junctions (each with inductance of 13\,nH). The on-chip features are fabricated on a sapphire substrate, which is then inserted into the high-purity (5N) aluminium package. The whole device is housed in a standard Cryoperm shield to protect the quantum circuit from external electromagnetic noise.

The system is thermally anchored to the base stage of the dilution refrigerator. To improve the thermalisation of the flux line, OFHC copper braids (as depicted in Fig.~\ref{fig:hose_chip}d) are used to help dissipate heat from the current applied to the coil. A short section of a copper braid connects the magnetic hose directly to the copper brackets that are affixed to the base plate of the fridge. Additionally, the two solder points along the line are directly thermally anchored to the base plate (one at the first eccosorb filter after the coil and another at the bias tee). With these procedures, we are able to access the full tunability range of the transmon without raising the mixing chamber temperature beyond 10\,mK.

The readout of the transmon state is performed with a standard reflection setup. The output signal is amplified by a phase-sensitive SNAIL Parametric Amplifier (SPA)~\cite{frattini_optimizing_2018, dorogov_application_2022} at the lower stage of the refrigerator. The signal is further amplified at the 4\,K stage by a low-noise HEMT amplifier (LNF-LNC4\_8C), and by a 26\,dB amplifier at room temperature (ZVA-183-S+) before demodulation.

\subsection{Room temperature configuration}

The transmon and cavity are controlled by a fast field-programmable gate array (FPGA) system from Quantum Machines (QM). The Digital-to-Analogue Converter (DAC) port of the FPGA is used to create signals within a 250\,MHz bandwidth. The control and probing pulses of the system are produced by upconverting intermediate frequency (IF) FPGA signals by IQ mixing with local oscillators (LO).

\begin{figure*}[p]
    \includegraphics[scale=1]{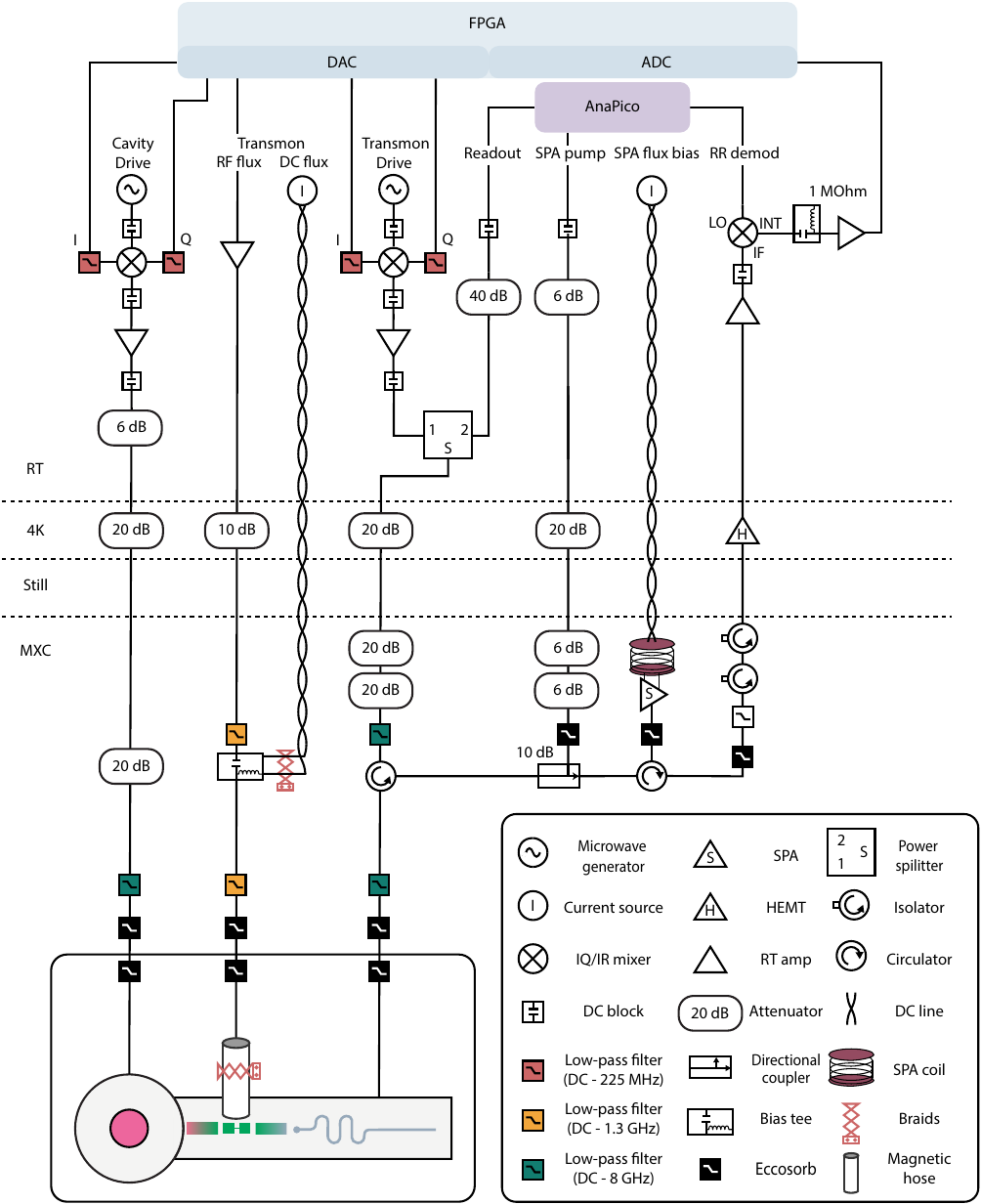}
    \caption{\textbf{Experimental setup.} The RF control lines are configured with a standard set of filters, including a low-pass filter (Mini-Circuits BLP-250+\,MHz) located outside the refrigerator and followed by a room-temperature amplifier Mini-Circuits ZVA-183-S+, working at GHz frequencies. Moreover, Eccosorb and standard low-pass filters (Hermerc System 8\,GHz) are also incorporated within the mixing chamber. The DC supplies to the coil are provided through built-in DC lines in the fridge, and combined with the RF drives via a cryogenic bias tee. }
    \label{fig: Wiring}
\end{figure*}

The DAC is also connected to an RF flux line that carries fast flux pulses to the coil. The coil is also biased with a stable current provided by a Yokogawa DC source, which is combined to the RF flux line using a bias tee (Marki Microwave BT0040) at the base plate of the fridge. For certain experiments, the fast flux signal is also amplified by a room-temperature amplifier (Stanford Research Systems SR445A) to provide a larger range of drive amplitudes to the coil. To mitigate high-frequency current noise at the coil, low-pass filters with 1.3\,GHz cutoff (Minicircuits VLFX-1300) are placed on the flux line. 

The readout tone, SPA pump, as well as an IF tone are all generated by a phase-locked multi-channel signal generator (AnaPico APMS-ULN) to ensure phase stability. The returning signal is downconverted through an IR mixer with an IF frequency of 250\,MHz. This specific frequency is chosen to align with the clock cycle of the FPGA, further ensuring that no additional phase is introduced between each readout cycle. The downconverted signal is subsequently sampled in an Analogue-to-Digital Converter (ADC) block of the FPGA. 

The schematic of the setup is summarised in Fig.~\ref{fig: Wiring}. 

\section{Flux pulse pre-distortion}
The transmon frequency $\omega_t(t)$ is controlled by the current $I_c(t)$ at the coil as a function of time, according to the relation
\begin{equation}
    \label{eq:flux_curve}
    \omega_t(t) \approx \left(\omega_t^{\mathrm{max}} + \alpha\right)\sqrt{\left|\cos\left(\frac{\pi kI_c}{\Phi_0}\right)\right|} - \alpha.
\end{equation}
The coil is driven through the RF line connected to the room-temperature electronics (Fig.~\ref{fig: Wiring}). The FPGA sends programmable flux pulses $V_{f}(t)$ to the line, which can be assumed to be linearly related to the coil current through the relation
\begin{equation}
    \label{eq:transferfn}
    \tilde{I}_c(s) = H_{\mathrm{line}}(s)\tilde{V}_{f}(s),
\end{equation}
where $\tilde{I}_c$ and $\tilde{V}_f$ are the Laplace representations of the current and voltage waveforms and $H_{\mathrm{line}}$ is a function that accounts for the linear response of the RF flux line.

The term, $H_{\mathrm{line}}$, includes the effective impedance of all components along the line, such as the bias tee, the filters, and the coil itself, leading to a complex distortion of the current waveform. These distortions have critical experimental consequences. Instability of the transmon frequency at the microsecond timescale causes imprecision in longer gates such as frequency-selective $\pi$-pulses. Also, the frequency transition speed can be slowed down by the presence of low-pass filtering, compromising experiments that require nonadiabatic frequency switching such as vacuum Rabi oscillations. The solution to accurately control of $\omega_t(t)$ is to program a predistorted flux pulse $V_f$ that takes the line response $H_{\mathrm{line}}$ into account and generates the intended $I_c$.

\begin{figure}[hbt!]
\includegraphics[scale=1]{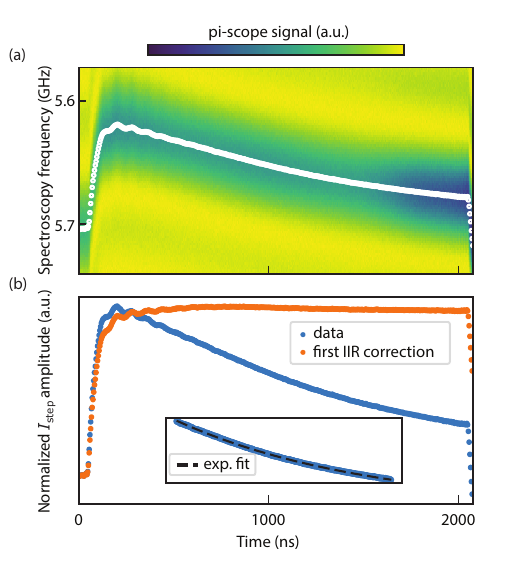}
\caption{\textbf{Flux pulse predistortion.} (a) Pi-scope experiment consisting of a time-dependent transmon spectroscopy during the application of a step flux pulse. $\omega_t(t)$ is extracted at each point of time (white dots). (b) Example of a first IIR correction. The current response is obtained from the transmon frequency data. An IIR filter is build from an exponential fit and used to correct the current response, making it more stable for longer periods of time.}
\label{fig:predistortion}
\end{figure}

To achieve the predistortion of the flux pulse, we first characterize the function $H_{\mathrm{line}}$ by measuring the current response to a step flux pulse $I_{\mathrm{step}}(t)$. The step response encodes all information about $H_{\mathrm{line}}$ and is obtained with a pi-scope experiment (Fig.~\ref{fig:predistortion}a). This method resolves the transmon frequency trajectory along the step flux pulse by executing multiple transmon spectroscopies as a function of time. The length of the spectroscopy probe pulse sets the time resolution of the experiment (we use 16\,ns constant pulses). The spectroscopy data is analyzed at each point of time to find the transmon frequency $\omega_t(t)$, which is used to obtain $I_{\mathrm{step}}(t)$ using Eq.~\ref{eq:flux_curve}.

From the step response, we train a set of digital infinite impulse response (IIR) and finite impulse response (FIR) filters that, combined, approximate the inverse function $H^{-1}_{\mathrm{line}}$~\cite{butscher_shaping_2018}. Any target current trajectory $I_{\mathrm{target}}$ can then be faithfully replicated by programming a predistorted flux pulse with the formula $\tilde{V}_f= H^{-1}_{\mathrm{line}}\tilde{I}_{\mathrm{target}}$. 

By successfully employing this method, we reach accurate control of the transmon trajectory, necessary for controlling the cavity-transmon interactions reliably. Here, we describe in more detail the formulas and the procedures involved in training the IIR and FIR filters.

\subsection{IIR filters}

Fig.~\ref{fig:predistortion}b shows $I_{\mathrm{step}}(t)$ obtained by measuring the $\omega_t(t)$ response to a step flux pulse with pi-scope and inverting Eq.~\ref{eq:flux_curve}. The trajectory shows a multi-exponentially decaying current,  along with a rising edge that also shows exponential behaviour. Our first distortion-correcting strategy is to build a set of IIR filters that are each tailored to invert a single exponential trend in the trajectory.

We can see how this works in a simplified example. Assume a step response of the form $I_{\mathrm{step}}=\left(A + Be^{-t/\tau}\right)u(t)$, where $u(t)$ is the Heaviside function. The corresponding transfer function can be calculated by taking the Laplace transform of $I_{\mathrm{step}}$ and using Eq.~\ref{eq:transferfn}
\begin{equation}
    H_{\mathrm{exp}} = \left(\frac{A}{s} + \frac{B\tau}{1+s\tau}\right)\tilde{V}_f^{-1},
\end{equation}
where $\tilde{V}_f = 1/s$ for the step response. The inverse of $H_{\mathrm{exp}}$ is simply
\begin{equation}
    H^{-1}_{\mathrm{exp}} = \frac{1 + s\tau}{A + s\tau\left(A + B\right)}.
\end{equation}
In practice, we want to build discrete-time predistortion filters that can be programmed in the FPGA. To discretize $H^{-1}_{\mathrm{exp}}$, we apply the bilinear transformation with a timestep $T_s$:
\begin{equation}
    s \leftarrow \frac{2}{T_s}\frac{1-z^{-1}}{1+z^{-1}}.
\end{equation}
This results in the digital predistortion IIR filter
\begin{equation}
    \label{eq:IIR_filter}
    H_{\mathrm{IIR}}(z) = \frac{b_0 + b_1z^{-1}}{1-a_1z^{-1}},
\end{equation}
where $a_1 = \left(2A\tau + 2B\tau - AT_s\right)/\lambda$, $b_0 = (2\tau + T_s)/\lambda$, $b_1 = (-2\tau + T_s)/\lambda$ and $\lambda = \left(2A\tau + 2B\tau + AT_s\right)$.

We correct $I_{\mathrm{step}}$ iteratively by fitting it to exponential functions to extract IIR filters according to Eq.~\ref{eq:IIR_filter}, as shown in Fig.~\ref{fig:predistortion}b. After the first filter $H_{\mathrm{IIR}, 1}$ is extracted, the $I_{\mathrm{step}}$ data is corrected before training the next pulse. This can be done by repeating the pi-scope experiment using a step input predistorted by $H_{\mathrm{IIR}, 1}$. Alternatively (as extracting a new pi-scope measurement is time-consuming), the expected correction to $I_{\mathrm{step}}$ can be calculated by numerically applying  $H_{\mathrm{IIR}, 1}$. This process is then repeated until there is no evident exponential trend in $I_{\mathrm{step}}$. 

Even after concluding the IIR corrections, $I_c$ might still present oscillatory behavior close to the pulse edges (such ripples can already be seen in Fig.~\ref{fig:predistortion}). Next, we discuss how these can be corrected with an FIR filter.

\subsection{FIR filter}

FIR filters are used to correct the fast ripples that remain after the IIR filter corrections. In the time-domain, Eq.~\ref{eq:transferfn} becomes the convolution
\begin{equation}
    \label{eq:FIRconvolution}
    I_c = h*V_f + n,
\end{equation}
where $h$ is the impulse response of the system, and $n$ is the noise present in the measurement of $I_c$. If $V_f$ is a step function, $h$ can be calculated using the step response $I_{\mathrm{step}}$ by 
\begin{equation}
    \label{eq:impulse_response}
    h = \frac{dI_{\mathrm{step}}}{dt}.
\end{equation}
As in the previous section, we want to find the inverse $h_{\mathrm{inv}}$. However, due to the presence of noise, it is not straightforward to invert the convolution in Eq.~\ref{eq:FIRconvolution}.

Instead, we explore the property that the convolution $h*h_{\mathrm{inv}} = \delta$, where $\delta$ is the Dirac delta (or Kronecker delta in the case of discrete-time functions). If the system input is $V_f = h_{\mathrm{inv}}$, then the output is $\delta$ up to measurement noise. So $h_{\mathrm{inv}}$ can be found numerically through the optimisation problem
\begin{equation}
    h_{\mathrm{inv}} = \underset{x}{\mathrm{argmin}}\{||h*x - \delta||\},
\end{equation}
where $||\cdot||$ represents the $\mathrm{L}_2$ norm. 

However, in this formulation, the solution $h_{\mathrm{inv}}$ will still be sensitive to the noise that $I_{\mathrm{step}}$ carries into $h$ from Eq.~\ref{eq:impulse_response}. To reduce the effect of noise, we use a cost function $\alpha||Dx||$:
\begin{equation}
    h_{\mathrm{inv}} = \underset{x}{\mathrm{argmin}}\{||h*x - \delta|| + \alpha||Dx||\},
\end{equation}
where $D$ is a differentiation operator and $\alpha$ is the weight of the cost function. This optimisation penalises fast-oscillating values of $x$ that result from over-fitting the noise.

Once the $h_{\mathrm{inv}}$ is known, it can be straightforwardly implemented with an FIR filter. In discrete-time formulation, an FIR filter has the general form
\begin{equation}
    y[n] = b*x = \sum_{i=0}^{N} b[n] x[n-i],
\end{equation}
where $b_i$ are the coefficients of the filter. By simply assigning $b[n] = h_{\mathrm{inv}}[n]$, the filter transforms a target pulse $x$ into the predistorted pulse $y$. Together with the previously trained IIR filters, this concludes the flux predistortion.

\section{Data Analysis}

\subsection{{Measurement and Post-Processing}}
\subsubsection{Wigner function tomography}
In the strong dispersive regime, we implement Wigner functions to the cavity state $\rho$ by performing a set of cavity displacements $\hat{D}(\beta)$ and measuring the photon-number parity $\hat{\mathcal{P}}$, which leads to
\begin{equation}
     W(\beta) = \frac{2}{\pi}\langle\hat{\mathcal{P}}\rangle =\frac{2}{\pi} \mathrm{Tr}(\hat{D}^{\dagger}(\beta) \rho \hat{D}(\beta)\hat{\mathcal{P}}).
\end{equation} 
The parity measurement involves interleaving a conditional cavity $\pi$-phase shift between two $R_y(\frac{\pi}{2})$ rotations, with $R_y(\frac{\pi}{2})$ representing a $\frac{\pi}{2}$ rotation around the y-axis as shown in Fig.~3b in the main text.

The cross-Kerr effect between the readout mode and the storage cavity introduces distortions to the readout signal when large displacements occur in the storage cavity. In order to mitigate this, and also to convert the readout signal to parity, we implemented the approach described in Ref.~\cite{sun_tracking_2014, vlastakis_deterministically_2013}. The idea is to perform two parity measurements for a vacuum state in the cavity, with the rotation axis of the second $\frac{\pi}{2}$ pulse to be $R_{y}(\frac{\pi}{2})$ and $R_{y}(-\frac{\pi}{2})$, respectively. These two measurements are one-dimensional cuts along $\mathrm{Im}[\beta]=0$. The difference between them corresponds to the parity value, which returns $\langle\hat{\mathcal{P}}\rangle = \pm 1$ for an ideal system with no state preparation and measurement errors. 

In practice, the measurement in our system has finite precision. The single-shot readout is performed using a SPA operating with 16\,dB of gain in a phase sensitive regime. We use a square readout pulse with a length of 400\,ns, and the signal is acquired over a duration of 1000\,ns. With additional FPGA delays, the total readout time extends to 1.5\,$\mu$s,  leading to a readout fidelity of  $1-[P(e|g) + P(g|e)]/2\approx 87.9\%$, where $P(i|j)$ is the probability of measuring the state $|i\rangle$ when having prepared the state $|j\rangle$.

%In Fig.~3 and 4a of the main text, a square readout pulse with a length of 400\,ns is utilized, and the signal is acquired over a duration of 1000\,ns. With additional FPGA delays, the total readout time extends to 1.5\,$\mu$s,  leading to a readout fidelity of  $1-[P(e|g) + P(g|e)]/2\approx 87.9\%$, where $P(i|j)$ is the probability of measuring the state $|i\rangle$ when having prepared the state $|j\rangle$. In Fig.~4b, two readout steps are involved. Hence, we reduced the readout duration to 252\,ns and acquiring the signal for 752\,ns. Considering additional FPGA delays, the total readout time is 1.252\,$\mu$s. The overall fidelity for the two readouts is approximately 87\%.

Furthermore, due to the presence of the cross-Kerr effect, a Gaussian curve is fitted to the data to obtain key calibration parameters, including amplitude and offset, based on the vacuum data. We then use the parameters extracted from this procedure to calibrate the Fock state Wigner tomography data shown in Fig.~3c of the main text.

\subsubsection{Characteristic function tomography}
In the weak dispersive regime, we employ characteristic functions, defined as ${C(\nu)=\mathrm{Tr}(\hat{D}(\nu) \rho)}$, to characterize the cavity state $\rho$. The characteristic function, as depicted in Fig.~3b bottom in the main text, is implemented using the same procedure as in Ref.~\cite{campagne_quantum_2020}, which consists of a transmon $\frac{\pi}{2}$ pulse followed by an echo conditional displacement (ECD) gate and another $\frac{\pi}{2}$ pulse. The ECD gate is composed of four cavity displacements and one transmon $\pi$ pulse. By choosing the phase of the second transmon $\frac{\pi}{2}$ pulse to be either $0^\circ$ or $90^\circ$, we can measure either the real part, $\mathrm{Re}[C(\nu)]$, or imaginary part, $\mathrm{Im}[C(\nu)]$, of the characteristic functions $C(\nu)$.

We use the vacuum state to calibrate the characteristic function measurement by sweeping the amplitude of the ECD gate. This is a one-dimensional measurement along $\mathrm{Im}[(\nu)]=0$ of the $\mathrm{Re}[C(\nu)]$ of the vacuum. The objective is to obtain a unit displacement amplitude with a targeted Gaussian standard deviation $\sigma = 1$. Using a Gaussian fit, we extract the main calibration parameters, including $\sigma$, offset, and amplitude, from the vacuum state and use them to calibrate the subsequent characteristic function measurements. In addition, we calibrate the reference frame of the transmon phase so the $\mathrm{Im}[C(\nu)]$ is null.

When measuring the characteristic function of relatively large coherent states ($|\alpha|\geq 2$), we observe an elevated background in certain regions of the phase space. We attribute these phenomena to the large displacements required to measure the characteristic function. When these displacements are along the original coherent state amplitude, we create a significant photon population in the cavity which likely causes other spurious dynamics in the system. We performed several control experiments to verify that this effect does not depend on the specific transmon pulse-sequence or the flux-pulses. Furthermore, it does not affect the features in the centre of the phase-space, which is the region where the key information about the coherent states is contained. Thus, in the final data in Fig.~4 of the main text, we calibrate the background to remove these spurious artifacts.

\subsection{Fidelity estimation}
We estimate the fidelity of the Fock states prepared using the vacuum Rabi experiment by computing the parity of the measured Wigner function, which yields $\mathcal{F}_1\approx0.90$ and $\mathcal{F}_2\approx0.76$, respectively.

\begin{figure}[hbt!]
\includegraphics[scale=1]{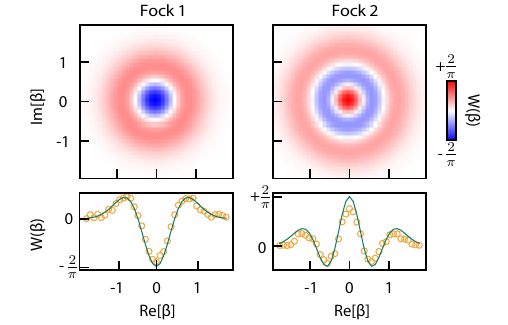}
\caption{\textbf{Comparison of simulated and measured Wigner functions of the Fock states.} Top: 2D plots of the simulated Fock $|1\rangle$ and $|2\rangle$, respectively. Bottom: line cuts along $\mathrm{Im}[\beta]=0$ in the simulated results (green line) and the corresponding measured data (orange cirles). The plots show a consistent agreement between the experimental data and the simulated behavior.}
\label{fig:wigner}
\end{figure}

We show the comparison between the 1D cuts in the ideal and measured Wigner function for the two states in Fig.~\ref{fig:wigner}. While the features of the states are in close agreement, there is a reduction in the overall contrast in the data, consistent with the estimated fidelities. We attribute this reduction to the imperfect preparation of $|e\rangle$ as the duration of the $\pi$-pulse used (80\,ns) is significant compared to the transmon $T_2$ ($\approx500$\,ns).

%By computing the Wigner function in the origin of the phase space of the Fock 1 and Fock 2 prepared using the vacuum Rabi experiment (shown in Fig.~3c in the main text), we get the value of -0.899 and 0.756, respectively. A comparison between the measured data and the simulated Wigner functions of the Fock states is drawn in the Fig.~\ref{fig:wigner}.

\subsection{Simulated cavity dynamics with different $\chi$ and $K$ configurations}  
\begin{figure}[bt!]
\includegraphics[scale=1]{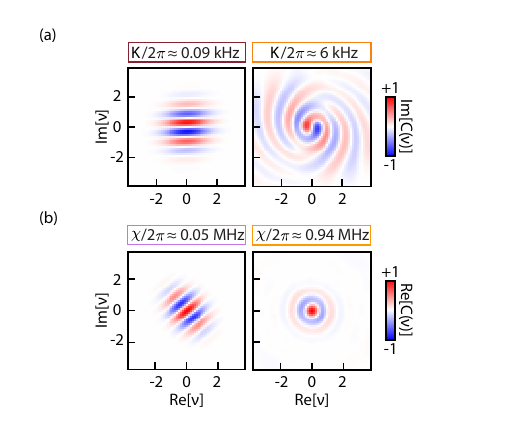}
\caption{\textbf{Simulation results of the suppression of undesired cavity dynamics.} (a) Mitigation of cavity self-Kerr. (b) Decoupling between transmon and cavity by turning off the dispersive interaction.}
\label{fig:kerr}
\end{figure}

In this section, we compare the experimental data presented in Fig.~4 in the main text with the corresponding simulation results. To model the effects of cavity self-Kerr and the decoupling between the transmon and the cavity, we use a standard master equation simulation that closely follows the experimental protocols outlined in the main text. The simulation also accounts for the independently calibrated parameters for cavity and transmon decoherence at the specific flux points used in the experiment. The resulting simulated state, shown in Fig.~\ref{fig:kerr}, is in close agreement with the data presented in the main text. This indicates that our experiment is faithfully reproducing the intended dynamics in the cavity. 

\subsection{Vacuum Rabi oscillations}  

\begin{figure}[bt!]
\includegraphics[scale=1]{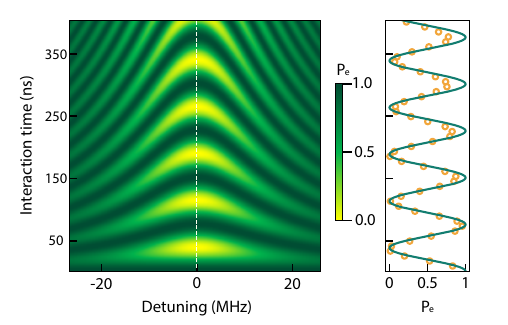}
\caption{\textbf{Simulation of vacuum Rabi oscillations between cavity and transmon.} Left: The coupling strength $g$ corresponds to the value found in experiment. Vacuum Rabi oscillations are shown as a function of interaction time and detuning, with the white dashed line denoting zero detuning. The colorbar indicates the population of transmon excited state. Right: A cut along the y-axis of simulated behaviour at zero detuning (green line) are compared to the experimental data (orange circles).}
\label{fig:vacuum Rabi}
\end{figure}

We directly compare the observed vacuum Rabi oscillations between the cavity and the transmon with numerical simulations using the measured coupling strength $g = 6.65$\,MHz. We observe close agreement between the simulation (shown in Fig.~\ref{fig:vacuum Rabi}) and the measurement results shown in Fig.~3a in the main text. 

The 1D cut along the y-axis shows the periodic oscillation that corresponds to the exchange of a single excitation between the transmon and the cavity. To account for the finite readout fidelity, we linearly transform the experimental points to map the highest observed transmon excited state population to 1 and the lowest to 0. The experimental results show an exponentially decaying oscillation due to the finite coherence times of the system. The amplitude of vacuum Rabi oscillation decays with a characteristic time of $\tau\approx1.5$\,$\mu$s, which is mainly limited by the decoherence time of the transmon but is increased due to hybridization with the cavity. 

\newpage
\bibliography{references.bib}    %use a bibtex bibliography file

\end{document}